\documentclass[twocolumn,pra,aps]{revtex4}

\usepackage{mathptmx}
\usepackage{subfigure}
\usepackage{psfrag,graphicx}
\usepackage{dcolumn}
\usepackage{amsmath,amssymb}
\usepackage{bm}
\usepackage{color}
\usepackage{latexsym}
\usepackage{epstopdf}
\usepackage{color}
\usepackage[english]{babel}
\usepackage{latexsym}
\usepackage{psfrag,graphicx}
\usepackage{subfigure}
\usepackage{amsmath}
\usepackage{amssymb}
\usepackage{amsfonts}
\usepackage{bm}
\usepackage{natbib}
\usepackage{epstopdf}
\usepackage{appendix}

\definecolor{mygrey}{gray}{0.35}
\definecolor{myblue}{rgb}{0.2,0.2,0.8}
\definecolor{myzard}{cmyk}{0,0,0.05,0}
\definecolor{mywhite}{rgb}{1,1,1}
\definecolor{mywhite}{rgb}{1,1,1}
\definecolor{myred}{rgb}{1,0.,0.3}

\usepackage[colorlinks=true,citecolor=myblue,linkcolor=myred]{hyperref}

\def\ba{\begin{align}}
\def\enda{\end{align}}
\def\bi{\begin{itemize}}
\def\ei{\end{itemize}}

\def\be{\begin{equation}}
\def\ee{\end{equation}}
\def\bea{\begin{eqnarray}}
\def\eea{\end{eqnarray}}
\def\bse{\begin{subequations}}
\def\ese{\end{subequations}}

\begin{document}
%\title{Circulant symmetry based on two-qubit quantum gate with trapped ions }
\title{Two-qubit quantum Fourier transform and entanglement protected by circulant symmetry }
\title{Two-qubit quantum gate and entanglement protected by circulant symmetry }
\author{Peter A. Ivanov}
\affiliation{Department of Physics, St. Kliment Ohridski University of Sofia, James Bourchier 5 blvd, 1164 Sofia, Bulgaria}
\author{Nikolay V. Vitanov}
\affiliation{Department of Physics, St. Kliment Ohridski University of Sofia, James Bourchier 5 blvd, 1164 Sofia, Bulgaria}

\begin{abstract}
We propose a method for the realization of the two-qubit quantum Fourier transform (QFT) using a Hamiltonian which possesses the circulant symmetry.
Importantly, the eigenvectors of the circulant matrices are the Fourier modes and do not depend on the magnitude of the Hamiltonian elements as long as the circulant symmetry is preserved.
The QFT implementation relies on the adiabatic transition from each of the spin product states to the respective quantum Fourier superposition states.
We show that in ion traps one can obtain a Hamiltonian with the circulant symmetry by tuning the spin-spin interaction between the trapped ions.
We present numerical results which demonstrate that very high fidelity can be obtained with realistic experimental resources.
We also describe how the gate can be accelerated by using a ``shortcut-to-adiabaticity'' field.
\end{abstract}

%\pacs{
%03.67.Ac, %Quantum computation architectures and implementations
%03.67.Bg,
%03.67.Lx,
%42.50.Dv %Coherent control of atomic interactions with photons
%}
\maketitle

%%%%%%%%%%%%%%%%%%%%%%%%%%%%%%%%%%%%%%%%%%%%%%%%%%%%%%%%%%%%%%%%%%%%%%%%%%%
%%%%%%%%%%%%%%%%%%%%%%%%%%%%%%%%%%%%%%%%%%%%%%%%%%%%%%%%%%%%%%%%%%%%%%%%%%%
%%%%%%%%%%%%%%%%%%%%%%%%%%%%%%%%%%%%%%%%%%%%%%%%%%%%%%%%%%%%%%%%%%%%%%%%%%%
%========================================================================
%========================================================================
\section{Introduction}

The quantum computers will dramatically accelerate particular computational tasks compared to the classical computers \cite{Ladd}.
Over the last 25 years, various quantum systems have been explored and used for the experimental realization of quantum computational tasks, including trapped ions \cite{Blatt2008}, trapped atoms \cite{Saffman2010}, photons \cite{Politi2009}, superconducting quantum circuits \cite{DiCarlo2009}, quantum dots \cite{Veldhorst2015}, doped solids \cite{Doherty2013}, etc.
Among these, superconducting qubits made recently headlines by demonstrating quantum supremacy \cite{Arute2019}, while trapped ions hold the records both in single-qubit \cite{Brown2011,Harty2014} and two-qubit gate fidelity \cite{Ballance2016,Gaebler2016}.
% have been reported in trapped-ions experiments.
%The unwanted cross-talk to neighboring qubits has been reduced to $10^{-5}$ \cite{Piltz2014} and $10^{-6}$ \cite{Craik2017} in other trapped-ions experiments, and ion transport with an error of less than $10^{-5}$ has been reported too \cite{Kaufmann2018}.
%Superconducting qubits have approached the gate fidelity benchmarks too \cite{Barends2014}.

The enabling condition for quantum computation is the ability to perform high-fidelity single- and two-qubit quantum gates.
Usually, the resonant quantum gates rely on an effective resonant interaction between the qubits which, however, makes the gate implementation sensitive to parameter fluctuations.
Alternatively, the quantum gates can be carried out by using adiabatic techniques, which are slower in time but more robust with respect to parameter fluctuations.

In this work, we propose an adiabatic implementation of the two-qubit quantum Fourier transform by using a Hamiltonian which possesses the circulant symmetry.
The unique property of the circulant matrices is that their eigenvectors are the Fourier modes.
Moreover, their eigenvectors do not depend on the magnitude of the Hamiltonian parameters as long as the circulant symmetry is preserved.
Such a circulant symmetry was studied as an efficient method for creation of superposition states in a single atom placed in a ring of quantum wells \cite{Unanyan2007},
as well as for implementation of a single-qubit gate \cite{Torosov2009}.
Here we consider a quantum system, which consists of two interacting spins in the presence of magnetic field.
We show that by proper adjustment of the spin-spin coupling and the single-qubit drive one can achieve a spin Hamiltonian with the circulant symmetry.
Our method relies on using adiabatic evolution which drives the system into the eigenstates of the circulant Hamiltonian and thereby realizes the quantum Fourier gate in a \emph{single} interaction step.
The adiabatic transition is performed by introducing a time-dependent energy offset of the spin states, which breaks the circulant symmetry but vanishes at the end of the transition.
We show that by a proper choice of the time-dependent couplings and detuning one can adiabatically transform any initial state into a superposition of quantum Fourier modes with high fidelity.

Since our technique relies on adiabatic evolution it is robust against parameter fluctuations and is mainly limited by the nonadiabatic transitions.
We show that for a specific choice of the parameters one can find exactly the eigenvectors of the full spin Hamiltonian at any instance of time.
This allows us to combine our gate scheme with the shortcuts to adiabaticity techniques \cite{Odelin2019} which can be used to suppress the effect of nonadiabatic transitions and thus to reduce the gate time.

We discuss the physical implementation of our gate scheme in a linear ion crystal driven by bichromatic laser fields.
Such an interaction creates a coupling between the internal states of the trapped ions with the collective vibrational modes.
We consider the dispersive regime in which the beatnote laser frequency is far off-resonant to any vibrational mode frequency.
In this regime the collective phonons can be traced out leading to an effective spin-spin interaction.
Such a regime where the phonons are only virtually excited was studied in the context of high-fidelity two-qubit gate implementation \cite{Kim2009,Bermudez2012,Tan2013}.
We show that by controlling the laser detuning we can perform the desired adiabatic evolution to the quantum Fourier modes.

The paper is organized as follows.
In Sec. \ref{model} we provide the general framework of the circulant-symmetric spin-spin Hamiltonian.
In Sec. \ref{adiabatic} we discuss the adiabatic transition to the quantum Fourier modes.
The physical realization of the circulant Hamiltonian using a laser driven ion crystal is discussed in Sec. \ref{PITI}.
In Sec. \ref{ne} we provide numerical estimation for the two-qubit gate fidelity as well as the fidelity for the creation of entangled states.
Finally, the conclusions are presented in Sec. \ref{C}.

\section{Model}\label{model}
We begin by considering two interacting spins which are subjected to a magnetic field.
The Hamiltonian of the system is given by
\begin{align}
\hat{H}&=J(\hat{\sigma}_{1}^{+}e^{-i\varphi_{1}}+\hat{\sigma}_{1}^{-}e^{i\varphi_{1}})(\hat{\sigma}_{2}^{+}e^{-i\varphi_{2}}+\hat{\sigma}_{2}^{-}e^{i\varphi_{2}})\notag\\
&+\Omega_{1}(\hat{\sigma}_{1}^{+}e^{i\phi_{1}}+\hat{\sigma}_{1}^{-}e^{-i\phi_{1}})+\Omega_{2}(\hat{\sigma}_{2}^{+}e^{i\phi_{2}}+\hat{\sigma}_{2}^{-}e^{-i\phi_{2}}),\label{Hc}
\end{align}
where $\sigma_{k}^{+}=\left|\uparrow_{k}\right\rangle\left\langle\downarrow_{k}\right|$ and $\sigma_{k}^{-}=\left|\downarrow_{k}\right\rangle\left\langle\uparrow_{k}\right|$ are the spin flip operators with $\left|\uparrow_{k}\right\rangle$ and $\left|\downarrow_{k}\right\rangle$ being the qubit states of the $k$th spin $(k=1,2)$.
The first term in (\ref{Hc}) describes the spin-spin interaction which is quantified by the coupling strength $J$ and phase $\varphi_{k}$.
The second and the third terms describe the single-qubit transitions with Rabi frequencies $\Omega_{1}$ and $\Omega_{2}$ and phases $\phi_1$ and $\phi_2$.
It is convenient to express the Hamiltonian in the computational basis formed by the qubit states $\{\left|\downarrow\downarrow\right\rangle,\left|\downarrow\uparrow\right\rangle,\left|\uparrow\downarrow\right\rangle,\left|\uparrow\uparrow
\right\rangle\}$.
Then the Hamiltonian becomes the 4$\times$4 hermitian matrix
\begin{equation}
H =\left[
\begin{array}{cccc}
0 & \Omega_{2}e^{-i\phi_{2}}&\Omega_{1} e^{-i\phi_{1}}& Je^{i(\varphi_{1}+\varphi_{2})} \\
\Omega_{2} e^{i\phi_{2}} & 0 &  Je^{-i(\varphi_{2}-\varphi_{1})} & \Omega_{1} e^{-i\phi_{1}} \\
\Omega_{1} e^{i\phi_{1}}&Je^{i(\varphi_{2}-\varphi_{1})}&0&\Omega_{2}e^{-i\phi_{2}}\\
J e^{-i(\varphi_{1}+\varphi_{2})}&\Omega_{1} e^{i\phi_{1}}&\Omega_{2}e^{i\phi_{2}}&0
\end{array}%
\right].\label{Hmatrix}
\end{equation}

In the following our goal is to find the conditions for the couplings $J$ and Rabi frequencies $\Omega_{1}$, $\Omega_{2}$ as well as for the phases $\phi_{k}$ and $\varphi_{k}$ such that the Hamiltonian (\ref{Hmatrix}) becomes a circulant matrix.
The important property of the circulant matrix is that its eigenvectors are the vector columns of the discrete quantum Fourier transform \cite{Gantmacher1986}.
Consequently, the eigenvectors do not depend on the elements of the circulant matrix but on the circulant symmetry only.
The most general 4$\times$4 circulant matrix has the following structure \cite{Gantmacher1986}:
\begin{equation}
C =\left[
\begin{array}{cccc}
c_{0}& c_{3}&c_{2}&c_{1} \\
c_{1}& c_{0} & c_{3} &c_{2} \\
c_{2}&c_{1}&c_{0}&c_{3}\\
c_{3}&c_{2}&c_{1}&c_{0}
\end{array}%
\right],\label{circulant}
\end{equation}
where $c_{p}$ ($p=0,\ldots,3$) are arbitrary complex numbers.
As can be seen the circulant matrix is completely defined by its first vector column (row) in the sense that all other columns (rows) are just cyclic permutations of it.
In the computational spin basis the eigenvectors of the 4$\times$4 circulant matrix can be expressed as
\bse\label{Fstates}
\begin{align}
|\psi_{0}\rangle &=\frac{1}{2}\{\left|\downarrow\downarrow\right\rangle+\left|\downarrow\uparrow\right\rangle+\left|\uparrow\downarrow\right\rangle+\left|\uparrow\uparrow\right\rangle\},\\
|\psi_{1}\rangle &=\frac{1}{2}\{\left|\downarrow\downarrow\right\rangle+i\left|\downarrow\uparrow\right\rangle-\left|\uparrow\downarrow\right\rangle-i\left|\uparrow\uparrow\right\rangle\},\\
|\psi_{2}\rangle &=\frac{1}{2}\{\left|\downarrow\downarrow\right\rangle-\left|\downarrow\uparrow\right\rangle+\left|\uparrow\downarrow\right\rangle-\left|\uparrow\uparrow\right\rangle\},\\
|\psi_{3}\rangle &=\frac{1}{2}\{\left|\downarrow\downarrow\right\rangle-i\left|\downarrow\uparrow\right\rangle-\left|\uparrow\downarrow\right\rangle+i\left|\uparrow\uparrow\right\rangle\}.
\end{align}
\ese
These four vectors (\ref{Fstates}) are the columns of the $4 \times 4$ quantum Fourier transform matrix.
Thus, by preparing the system in the eigenstates of the circulant Hamiltonian one can implement the two-qubit quantum Fourier transform.

In order to fulfill the circulant cyclic permutation symmetry we consider two different cases.

\emph{Case 1}: We have
\begin{equation}
J=\Omega_{2},\quad \Omega_{1}=0,\quad \varphi_{2}=\phi_{2}=\varphi,\quad \varphi_{1}=2p\pi,\label{cond1}
\end{equation}
with $p$ being integer.
The first condition requires the spin-spin coupling to be equal to the Rabi frequency on the second spin.
The circulant symmetry of the Hamiltonian (\ref{Hmatrix})  leaves arbitrariness in the choice of the Rabi frequency on the first spin.
Here we have set to zero, $\Omega_{1}=0$.
Using this, the Hamiltonian \eqref{Hmatrix} becomes a circulant matrix and can be rewritten as
\begin{equation}
\hat{H}_{\rm cir}^{(1)}=J(\hat{\sigma}_{1}^{+}+\hat{\sigma}_{1}^{-})(\hat{\sigma}_{2}^{+}e^{-i\varphi}+\hat{\sigma}_{2}^{-}e^{i\varphi})+J (\hat{\sigma}_{2}^{+}e^{i\varphi}+\hat{\sigma}_{2}^{-}e^{-i\varphi}).\label{H1}
\end{equation}

\emph{Case 2}:
The same conditions as (\ref{cond1}) but now with
\begin{equation}
\Omega_{1}\neq0,\quad \phi_{1}=p\pi.\label{cond2}
\end{equation}
Again the Hamiltonian is circulant and can be expressed as
\begin{eqnarray}
\hat{H}_{\rm cir}^{(2)}&=&J(\hat{\sigma}_{1}^{+}+\hat{\sigma}_{1}^{-})(\hat{\sigma}_{2}^{+}e^{-i\varphi}+\hat{\sigma}_{2}^{-}e^{i\varphi})+J (\hat{\sigma}_{2}^{+}e^{i\varphi}+\hat{\sigma}_{2}^{-}e^{-i\varphi})\notag\\
&&+\Omega_{1}(\hat{\sigma}_{1}^{+}+\hat{\sigma}_{1}^{-}).\label{H2}
\end{eqnarray}
We will show latter on that the additional Rabi frequency $\Omega_{1}$ in the circulant Hamiltonian in \emph{Case 2} can be used to improve significantly the adiabatic evolution even when the spin-spin coupling $J$ is rather small.
%of order of few kHz.

\section{Adiabatic Transition to Fourier Modes}\label{adiabatic}

In order to implement the two-qubit Fourier transform we assume that additionally to the circulant Hamiltonian time-dependent frequency shifts are applied such that the total Hamiltonian becomes
\begin{equation}
\hat{H}(t) =\hat{H}_{0}(t)+\hat{H}_{\rm cir}^{(j)}(t),\label{Htotal}
\end{equation}
with
\begin{equation}
\hat{H}_{0}(t) =\Delta_{1}(t)\hat{\sigma}_{1}^{z}+\Delta_{2}(t)\hat{\sigma}_{2}^{z},\label{H0}
\end{equation}
where $\Delta_{k}(t)$ is the time-dependent detuning of the $k$th spin.
Such a term is needed to control the adiabatic transition of the computational spin states to the quantum Fourier states (\ref{Fstates}).

Let us assume that initially the system is prepared in one of the computational product states $|\psi_{s_{1}s_{2}}\rangle=|s_{1}s_{2}\rangle$ ($s_{k}=\downarrow_{k},\uparrow_{k}$) which is an eigenstate of the Hamiltonian $\hat{H}_{0}(t)$.
%We assume that initially the system is prepared in an eigenstate of $\hat{H}(t_{\rm i})$.
As long as at the initial moment $t_{\rm i}$ the detuning $\Delta_{1,2}(t_{\rm i})$ is mich higher than the couplings $ J(t_{i}), \Omega_{1}(t_{i})$, i.e. $\Delta_{1,2}(t_{\rm i})\gg J(t_{i}), \Omega_{1}(t_{i})$ the respective eigenstates of the Hamiltonian (\ref{Htotal}) coincide with the computational spin states, namely
$|\psi(t_{\rm i})\rangle=|\psi_{s_{1}s_{2}}\rangle$.
Then we adiabatically decrease in time the detunings $\Delta_{1}(t)$ and $\Delta_{2}(t)$ to zero, while we increase the couplings $J(t)$ and $\Omega_{1}(t)$ such that in the end we have $\hat{H}_{0}(t_f) \to 0$ and $\hat{H}(t) \to \hat{H}_{\rm cir}^{(j)}(t)$.
In the adiabatic limit, the system remains in the same eigenstate of the full Hamiltonian $\hat{H}(t)$ at all times.
With the chosen time behavior of the couplings and the detunings, each such eigenstate is equal to a computational spin state (eigenstate of $\hat{H}_{0}$) in the beginning,  $|\psi(t_{\rm i})\rangle=|\psi_{s_{1}s_{2}}\rangle$, and to a Fourier state (eigenstate of $\hat{H}_{\rm cir}^{(j)}(t)$) in the end, $|\psi(t_{f})\rangle=|\psi_{p}\rangle$ ($p=0,1,2,3$).
Hence the adiabatic evolution maps each computational spin state onto a Fourier state, thereby producing the quantum Fourier transform in a single interaction step.

The adiabatic evolution requires that the separation between the eigenefrequencies $\lambda_{\pm}^{(j)}$ and $\mu_{\pm}^{(j)}$ of $\hat{H}(t)$ is larger at any instance of time than the nonadiabatic coupling between each pair of the eigenstates $|\lambda_{\pm}^{(j)}\rangle$ and $|\mu_{\pm}^{(j)}\rangle$ of $\hat{H}(t)$, i.e.
\begin{subequations}
\begin{eqnarray}
&&|\lambda_{\pm}^{(j)}(t)-\mu_{\pm}^{(j)}(t)|\gg |\langle\partial_{t}\lambda_{\pm}^{(j)}(t)|\mu_{\pm}^{(j)}(t)\rangle|,\\
&&|\mu_{+}^{(j)}(t)-\mu_{-}^{(j)}(t)|\gg |\langle\partial_{t}\mu_{+}^{(j)}(t)|\mu_{-}^{(j)}(t)\rangle|.
\end{eqnarray}
\end{subequations}
For smoothly varying Hamiltonian parameters adiabatic evolution usually demands that the interaction duration $T$ is large compared to the inverse of the smallest coupling or detuning implying large pulse areas and/or large detuning areas.

\subsection{Case 1}

Let us consider the eigenspectrum of the total Hamiltonian \eqref{Htotal}.
Consider first the circulant Hamiltonian $H_{\rm cir}^{(1)}$, i.e. $\hat{H}(t) = \hat{H}_{0}(t) + H_{\rm cir}^{(1)}$.
We find that the eigenfrequencies of $\hat{H}(t)$ are
\bse\label{ef}
\begin{align}
\lambda^{(1)}_{\pm}&=\pm\{2J^{2}+\Delta_{1}^{2}+\Delta_{2}^{2}
+2[J^{4}\cos^{2}2\varphi +\Delta_{1}^{2}(J^{2}+\Delta_{2}^{2})]^{\frac12}\}^{\frac12}, \\
%\notag\\ &+2[J^{4}\cos^{2}(2\varphi)+\Delta_{1}^{2}(J^{2}+\Delta_{2}^{2})]^{1/2}\}^{1/2}, \\
\mu^{(1)}_{\pm}&= \pm\{2J^{2}+\Delta_{1}^{2}+\Delta_{2}^{2}
-2[J^{4}\cos^{2}2\varphi +\Delta_{1}^{2}(J^{2}+\Delta_{2}^{2})]^{\frac12}\}^{\frac12},
%\notag\\ &-2[J^{4}\cos^{2}(2\varphi)+\Delta_{1}^{2}(J^{2}+\Delta_{2}^{2})]^{1/2}\}^{1/2},
\end{align}
\ese
which correspond to the eigenvectors $|\lambda_{\pm}^{(1)}\rangle$ and $|\mu_{\pm}^{(1)}\rangle$.
Note that in order to drive the adiabatic transition we require that the eigenfrequencies are nondegenerate at any instance of time.
Otherwise the system may evolve into a superposition of Fourier states which will spoil the gate implementation.
Initially  we begin with $\Delta_{1,2}(t_{\rm i})\gg J(t_{\rm i})$, such that the eigenfrequencies are $\lambda^{(1)}_{\pm}(t_{\rm i})=\pm(\Delta_{1}(t_{\rm i})+\Delta_{2}(t_{\rm i}))$ and respectively, $\mu^{(1)}_{\pm}(t_{\rm i})=\pm(\Delta_{1}(t_{\rm i})-\Delta_{2}(t_{\rm i}))$.
Hence in order to have nondegenerate spectrum we require $\Delta_{1}(t_{\rm i})\neq\Delta_{2}(t_{\rm i})$.
The eigenfrequencies will be equidistant if $\Delta_{1}(t_{\rm i})/\Delta_{2}(t_{\rm i}) = \frac13$ or $3$.

At the final instance of time where $\Delta_{1,2}(t_{\rm f})\ll J(t_{\rm f})$ the Hamiltonian possesses circulant symmetry.
At this final stage of the adiabatic transition the eigenfrequencies becomes
\begin{equation}
\lambda^{(1)}_{\pm}(t_{\rm f})=\pm 2J(t_{\rm f})\cos(\varphi),\quad \mu^{(1)}_{\pm}(t_{\rm f})=\pm 2J(t_{\rm f})\sin(\varphi),\label{cr1}
\end{equation}
with corresponding eigenvectors $|\lambda_{+}^{(1)}\rangle=|\psi_{0}\rangle$, $|\lambda_{-}^{(1)}\rangle=|\psi_{2}\rangle$, $|\mu_{+}^{(1)}\rangle=|\psi_{1}\rangle$ and $|\mu_{-}^{(1)}\rangle=|\psi_{3}\rangle$.
As can be seen there exist a finite energy gap for any phase $\varphi$ except for $\varphi=n\pi/4$, ($n=0,\pm1,\pm2,\ldots$) where the spectrum becomes degenerate.
The gaps are equal when $\tan(\varphi)= \frac13$ or $3$, i.e. when $\varphi = \arctan(3) \approx 0.3976\pi$ or $\varphi = \arctan(1/3) \approx 0.1024\pi$.

To summarize, the conditions for the scheme to work in this case are
\begin{subequations}
\begin{align}
\Delta_{1}(t_{\rm i}) &\neq\Delta_{2}(t_{\rm i}) , \\
\varphi &\neq  n\pi/4 \quad (n\ \text{integer}).
\end{align}
\end{subequations}

\subsection{Case 2}

Alternatively, one can drive the adiabatic transition to the Fourier states using the circulant Hamiltonian $\hat{H}_{\rm cir}^{(2)}$, i.e. $\hat{H}(t) = \hat{H}_{0}(t) + H_{\rm cir}^{(2)}$.
In order to get insight of the eigenfrequencies we set the phase to $\varphi=\pi/4$ which allows analytical treatment. We find
\bse\label{ef1}
\begin{align}
\lambda^{(2)}_{\pm}&= \pm\{2J^{2}+\Omega_{1}^{2}+\Delta_{1}^{2}+\Delta_{2}^{2}\notag\\
&+2[J^{2}(2\Omega_{1}^{2}+\Delta_{1}^{2})+\Delta_{2}^{2}(\Omega_{1}^{2}+\Delta_{1}^{2})]^{\frac12}\}^{\frac12}, \\
\mu^{(2)}_{\pm}&= \pm\{2J^{2}+\Omega_{1}^{2}+\Delta_{1}^{2}+\Delta_{2}^{2}\notag\\
&-2[J^{2}(2\Omega_{1}^{2}+\Delta_{1}^{2})+\Delta_{2}^{2}(\Omega_{1}^{2}+\Delta_{1}^{2})]^{\frac12}\}^{\frac12}.
\end{align}
\ese
We denote the corresponding instantaneous eigenvectors by $|\lambda_{\pm}^{(2)}\rangle$ and $|\mu_{\pm}^{(2)}\rangle$.
Again initially we start with $\Delta_{1,2}(t_{\rm i})\gg J(t_{\rm i}),\Omega_{1}(t_{\rm i})$ which indicates that $\lambda^{(2)}_{\pm}(t_{\rm i})=\pm(\Delta_{1}(t_{\rm i})+\Delta_{2}(t_{\rm i}))$ and respectively $\mu^{(2)}_{\pm}(t_{\rm i})=\pm(\Delta_{1}(t_{\rm i})-\Delta_{2}(t_{\rm i}))$.
As in \emph{Case 1}, the condition $\Delta_{1}(t_{\rm i})\neq\Delta_{2}(t_{\rm i})$ must be fulfilled in order to avoid degeneracy.
Equidistant eigenfrequencies occur initially if $\Delta_{1}(t_{\rm i}) = 3\Delta_{2}(t_{\rm i})$ or $\Delta_{2}(t_{\rm i}) = 3\Delta_{1}(t_{\rm i})$.
In the end, $J(t_{\rm f}),\Omega_{1}(t_{\rm f})\gg\Delta_{1,2}(t_{\rm f})$, the system arrives in an eigenstate of the circulant Hamiltonian $H_{\rm cir}^{(2)}$.
For any value of $\varphi$ the circulant eigenfrequencies at $t_{\rm f}$ are given by
\begin{eqnarray}
&&\lambda^{(2)}_{+}(t_{\rm f})=\Omega_{1}+2J\cos(\varphi),\quad \lambda^{(2)}_{-}(t_{\rm f})=-\Omega_{1}-2J\sin(\varphi),\notag\\
&&\mu^{(2)}_{+}(t_{\rm f})=\Omega_{1}-2J\cos(\varphi),\quad \mu^{(2)}_{-}(t_{\rm f})=-\Omega_{1}+2J\sin(\varphi),\label{eigenfrequencies2}
\end{eqnarray}
with corresponding eigenvectors $|\lambda_{+}^{(2)}\rangle=\left|\psi_{0}\right\rangle$, $|\lambda^{(2)}_{-}\rangle=\left|\psi_{3}\right\rangle$, $|\mu^{(2)}_{+}\rangle=|\psi_{1}\rangle$, and $|\mu^{(2)}_{-}\rangle=|\psi_{2}\rangle$. Assuming that $\Omega_{1}\neq 2J$ we see that the spectrum is nondegenerate except for $\varphi=n\pi/2$ with $n$ being integer.

\subsection{Transitions}

Let us now discuss the set of transitions which realize the quantum Fourier transform.
For concreteness we focus on the case with $\Omega_{1}\neq 0$ and choose the phase $\varphi=\pi/4$, with eigenfrequencies (\ref{ef1}).
Initially, each of the computational spin states coincide with the eigenvectors of the Hamiltonian (\ref{Htotal}), namely $|\lambda^{(2)}_{+}\rangle=\left|\uparrow\uparrow\right\rangle$, $|\lambda^{(2)}_{-}\rangle=\left|\downarrow\downarrow\right\rangle$, and $|\mu^{(2)}_{+}\rangle=\left|\uparrow\downarrow\right\rangle$, $|\mu^{(2)}_{-}\rangle=\left|\downarrow\uparrow\right\rangle$.
The realization of the quantum Fourier transform relies on the adiabatic following of each of the instantaneous eigenvectors,
%Thereby, one can realize the following set of transitions:
\bse \label{transition}
\begin{align}
&\left|\downarrow\downarrow\right\rangle\rightarrow e^{i\alpha_{2}}\left|\psi_{3}\right\rangle, \\
&\left|\downarrow\uparrow\right\rangle\rightarrow -ie^{i\beta_{2}}|\psi_{1}\rangle, \\
&\left|\uparrow\downarrow\right\rangle\rightarrow e^{-i\beta_{2}}|\psi_{2}\rangle, \\
&\left|\uparrow\uparrow\right\rangle\rightarrow e^{-i\alpha_{2}}\left|\psi_{0}\right\rangle.
\end{align}
\ese
Here $\alpha_{2}=\int_{t_{\rm i}}^{t_{\rm f}}\lambda_{+}^{(2)}(t)dt$ and $\beta_{2}=\int_{t_{\rm i}}^{t_{\rm f}}\mu_{+}^{(2)}(t)dt$ are the global adiabatic phases which appear due to the adiabatic evolution.
As we will show latter on by a proper choice of the detunings $\Delta_{1,2}$ the adiabatic phases can be tuned to be $\alpha_{2}=2p\pi$ and $\beta_{2}=2m\pi$ with $p$ and $m$ being integers.
This choice realises the following gate
\begin{equation}
G_{\frac{\pi}{4}} =\frac{1}{2}\left[
\begin{array}{cccc}
1& -i&1&1 \\
-i& 1 & -1 &1 \\
-1&i&1&1\\
i&-1&-1&1
\end{array}%
\right].\label{gate}
\end{equation}
Up to an additional phase factor $-\pi/2$ in the second column, the matrix \eqref{gate} resembles the quantum Fourier transform for two qubits.
This phase factor appears due to the determinant invariance during the adiabatic evolution, which imposes the requirement $\det G_{\frac{\pi}{4}}=1$.

Finally, we point out that if we replace $\varphi=\pi/4$ by $\varphi=-\pi/4$ then two of the circulant eigenfrequencies (\ref{eigenfrequencies2}) interchange $\lambda_{-}^{(2)}\leftrightarrow\mu_{-}^{(2)}$ and hence the adiabatic following of the eigenstates implies that $\left|\downarrow\downarrow\right\rangle\rightarrow e^{i\alpha_{2}}|\psi_{1}\rangle$ and $\left|\downarrow\uparrow\right\rangle\rightarrow i e^{i\beta_{2}}|\psi_{3}\rangle$. Hence the unitary matrix for this case becomes $G_{-\frac{\pi}{4}}=(G_{\frac{\pi}{4}})^{*}$.

\begin{figure}
\includegraphics[width=0.45\textwidth]{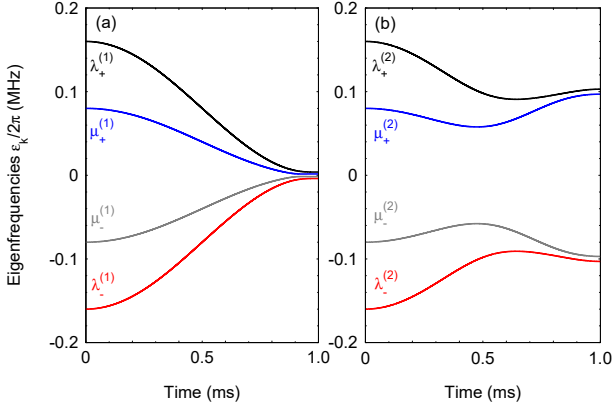}
\caption{(Color online) (a) Eigenfrequencies (\ref{ef}) of the Hamiltonian $\hat{H}(t)=\hat{H}_{0}(t)+\hat{H}_{\rm cir}^{(1)}(t)$ versus the interaction time. The coupling and the detuning vary in time according Eq. (\ref{parameters}). The parameters are set to $\Delta_{1}/2\pi=120$ kHz, $\Delta_{2}/2\pi=30$ kHz, $J_{0}/2\pi=2.1$ kHz, $\varphi=\pi/8$, and $\omega/2\pi=0.25$ kHz. (b) The same but for the Hamiltonian $\hat{H}(t)=\hat{H}_{0}(t)+\hat{H}_{\rm cir}^{(2)}(t)$. The parameters are set to $\Omega_{1}/2\pi=100$ kHz and  $\varphi=\pi/4$.}
\label{fig1}
\end{figure}
\subsection{Controlling the single qubit Rabi frequency}

The transition to the circulant Hamiltonian (\ref{H2}) (\emph{Case 2}) can be carried out even without the presence of energy offset described by Eq. (\ref{H0}). Indeed, let's set the phases in (\ref{Hc}) to $\varphi_{2}=\phi_{2}=\varphi$ and $\phi_{1}=2p\pi$. Then we have
\begin{align}
\hat{H}&=J(\hat{\sigma}_{1}^{+}+\hat{\sigma}_{1}^{-})(\hat{\sigma}_{2}^{+}e^{-i\varphi}+\hat{\sigma}_{2}^{-}e^{i\varphi})\notag\\
&+\Omega_{1}(\hat{\sigma}_{1}^{+}+\hat{\sigma}_{1}^{-})+\Omega_{2}(\hat{\sigma}_{2}^{+}e^{i\varphi}+\hat{\sigma}_{2}^{-}e^{-i\varphi}).\label{H3}
\end{align}
The Hamiltonian (\ref{H3}) has no circulant symmetry because the condition $J=\Omega_{2}$ is not fulfilled. However, the adiabatic transition to the Fourier modes can be carried out for example by varying in time the Rabi frequency $\Omega_{2}(t)$. At the initial moment we begin with $\Omega_{1},\Omega_{2}(t_{\rm i})\gg J$ such that the eigenstates are $|\psi(t_{\rm i})\rangle=|q_{1}q_{2}\rangle$, ($q_{k}=\pm$) where $|\pm_{1}\rangle=(\left|\downarrow_{1}\right\rangle\pm\left|\uparrow_{1}\right\rangle)/\sqrt{2}$ and $|\pm_{2}\rangle=(\left|\downarrow_{2}\right\rangle\pm e^{i\varphi}\left|\uparrow_{2}\right\rangle)/\sqrt{2}$. Then, adiabatically decrease $\Omega_{2}(t)$ such that at the final instance of time we have $\Omega_{2}(t_{\rm f})=J$. Adiabatically following the instantaneous eigenstates transform the initial states into the respective quantum Fourier states (see the Supplement for the derivation). In contrast to the gate realization with nonzero detuning, now the adiabatic transition is carried out between the initial rotating computation spin states and the quantum Fourier states. Finally, we point out that instantaneous eigenvectors of Hamiltonian (\ref{H3}) can be found exactly, which allows to combine the gate scheme with the shortcuts to adiabaticity technique (see the Supplement for more details).

\section{Physical Implementation with Trapped Ions}\label{PITI}

The implementation of our gate scheme can be realized in various quantum optical systems, for example, including superconductiong qubits coupled to transmission lines \cite{Gu2017}, as well as using color center in nanodiamonds coupled to carbon nanotubes \cite{Li2016}.
Here we consider a trapped-ion realization of the circulant Hamiltonian.
Consider a linear ion crystal which consists of $N$ ions with mass $M$, aligned along the trap axis $z$ with radial and axial trap frequencies $\omega_{x}$, $\omega_{z}$.
The qubit system typically consists of two metastable levels $\left|\uparrow\right\rangle$, $\left|\downarrow\right\rangle$ of the trapped ion with energy difference $\omega_{0}$.
The small radial vibrations around the equilibrium positions are described by a set of collective vibrational modes with a Hamiltonian $\hat{H}_{\rm ph}=\sum_{n}\omega_{n}\hat{a}_{n}^{\dag}\hat{a}_{n}$ \cite{Schneider2012}.
Here $\hat{a}^{\dag}_{n}$, $\hat{a}_{n}$ are the phonon creation and annihilation operators of the $n$th vibrational mode with a frequency $\omega_{n}$.
Including the internal energy of the qubits $\hat{H}_{\rm q}=\sum_{k}\omega_{0}\sigma_{k}^{z}/2$ the interaction-free Hamiltonian becomes $\hat{H}_{0}=\hat{H}_{\rm q}+\hat{H}_{\rm ph}$.

\begin{figure}
\includegraphics[width=0.45\textwidth]{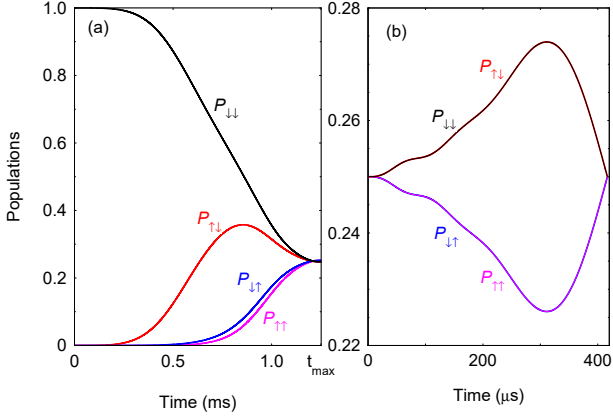}
\caption{(Color online) (a) Time evolution of the spin populations. We solve numerically the time-dependent Schr\"odinger equation with Hamiltonian $\hat{H}(t)=\hat{H}_{0}+\hat{H}_{\rm cir}^{(2)}$. The parameters are set to $J_{0}/2\pi=2$ kHz, $\Omega_{1}/2\pi=50$ kHz, $\Delta_{1}/2\pi=30$ kHz, $\Delta_{2}/2\pi=10$ kHz, $\omega/2\pi=0.2$ kHz, $\varphi=\pi/4$ and gate time $t_{\rm max}=1.25$ ms. (b) Adiabatic transition to the quantum Fourier state $\left|\psi_{3}\right\rangle$ using Hamiltonian (\ref{H3}). The parameters are set to $J_{0}/2\pi=2.0$ kHz, $V_{0}/2\pi=3.8$ kHz, $\Omega_{1}/2\pi=30$ kHz, $\omega/2\pi=0.6$ kHz, $\varphi=\pi/4$ and gate time $t_{\rm max}\approx 417$ $\mu$s.}
\label{fig2}
\end{figure}

In order to induce an effective spin-spin interaction between spin states we assume that an optical spin-dependent force is applied which couples the internal states of the ions with the collective vibrational modes \cite{Wineland1998,Haffner2008,Lee2005}.
In the following we assume that the desired spin-spin interaction is mediated by the radial phonons which are less sensitive to ion heating and thermal motion \cite{Zhu2006}.
Consider that each ion interacts with two pairs of noncopropagating laser beams along the radial direction with laser frequencies $\omega_{k,\rm L_{\rm r}}=\omega_{0}-\mu-\delta_{k}(t)$, $\omega_{k,\rm L_{\rm b}}=\omega_{0}+\mu-\delta_{k}(t)$ which give rise to a spin dependent force at frequency $\mu$. Here $\delta_{k}(t)=\int_{0}^{t}\Delta_{k}(\tau)d\tau$ is the small time-dependent laser detuning ($\omega_{0},\mu\gg\delta_{k}(t)$) of the ac Stark shifted states with respect to $\omega_{0}$ which introduce an effective qubit frequency.
In order to induce a single-spin transition we assume that the each ion interacts with a pair of copropagating laser beams with a frequency difference $\omega_{k,\rm L} =\omega_{0}-\delta(t)$.
Assuming the optical rotating-wave approximation (RWA) the interaction Hamiltonian becomes \cite{Wineland1998}
\begin{align}
\hat{H}_{\rm I}&= \sum_{k}\Delta_{k}\sigma_{k}^{z}+\Omega_{x}\sum_{k}e^{i k \hat{x}_{k}}\cos(\mu t)(e^{i\varphi_{k}}\sigma_{k}^{+}+e^{-i\varphi_{k}}\sigma_{k}^{-})\notag\\
&+\sum_{k}\Omega_{k}(e^{-i\phi_{k}}\sigma_{k}^{+}+e^{i\phi_{k}}\sigma_{k}^{-}).
\end{align}
Here $\Omega_{x}$, $\Omega_{k}$ are the Rabi frequencies, and respectively, $\varphi_{k}$, $\phi_{k}$ are the laser phases.
The small radial oscillations of the $k$th ion can be written in terms of collective normal modes, $k \hat{x}_{k}=\sum_{n}\eta_{k,n}(\hat{a}_{n}^{\dag}e^{i\omega_{n}t}+\hat{a}_{n}e^{-i\omega_{n}t})$, where $\eta_{k,n}=b_{k,n}k\sqrt{\hbar/2M\omega_{n}}$ is the Lamb-Dicke parameters with $b_{k,n}$ being the normal mode transformation matrix for the $k$ ion.
Within the Lamb-Dicke regime where $\Delta k \langle x_{k}\rangle\ll 1$ and performing the vibrational RWA we arrive at
\begin{align}
\hat{H}_{\rm I}&=\sum_{k}\Delta_{k}\sigma_{k}^{z}+\sum_{k,n}g_{k,n}\cos(\mu t)(\sigma_{k}^{+}e^{i\varphi_{k}}+\sigma_{k}^{-}e^{-i\varphi_{k}})\notag\\
&\times(\hat{a}^{\dag}_{n}e^{i\omega_{n}t}+\hat{a}_{n}e^{-i\omega_{n}t})+\sum_{k}\Omega_{k}(e^{-i\phi_{k}}\sigma_{k}^{+}+e^{i\phi_{k}}\sigma_{k}^{-}),
\end{align}
where $g_{k,n}=\eta_{k,n}\Omega_{x}$ is the spin-phonon coupling.
\begin{figure}
\includegraphics[width=0.45\textwidth]{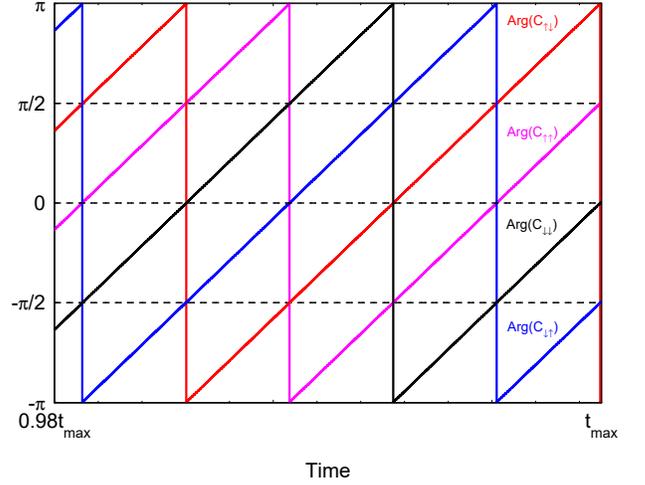}
\caption{(Color online) (a) Time evolution of the arguments of the probability amplitudes $C_{s_{1}s_{2}}(t)$. We solve numerically the time-dependent Schr\"odinger equation with Hamiltonian $\hat{H}(t)=\hat{H}_{0}+\hat{H}_{\rm cir}^{(2)}$ assuming the initial condition $|\psi(0)\rangle=e^{-i\alpha_{2}}\left|\downarrow\downarrow\right\rangle$. The parameters are set to $J_{0}/2\pi=2$ kHz, $\Omega_{1}/2\pi=50$ kHz, $\Delta_{1}/2\pi=30$ kHz, $\Delta_{2}/2\pi=10$ kHz, $\omega/2\pi=0.2$ kHz, $\varphi=\pi/4$.}
\label{fig3}
\end{figure}

We consider the regime in which the beatnote frequency $\mu$ is not resonant with any radial vibration mode and the condition $|\omega_{n}-\mu|\gg g_{k,n}$ is satisfied for any mode $n$.
In that case the radial collective phonons are only virtually excited, thereby they can be eliminated from the dynamics \cite{James2007}.
As a result of that the ion's spin states at different sites become coupled.
Finally, by assuming that only the $k$th and $m$th ions interact with the bichromatic field we obtain
\begin{align}
\hat{H}_{\rm I}&=J(\sigma_{k}^{+}e^{i\varphi_{k}}+\sigma_{k}^{-}e^{-i\varphi_{k}})(\sigma_{m}^{+}e^{i\varphi_{m}}+\sigma_{m}^{-}e^{-i\varphi_{m}})\notag\\
&+\Omega_{k}(e^{-i\phi_{k}}\sigma_{k}^{+}+e^{i\phi_{k}}\sigma_{k}^{-})+\Omega_{m}(e^{-i\phi_{m}}\sigma_{m}^{+}+e^{i\phi_{m}}\sigma_{m}^{-})\notag\\
&+\Delta_{k}\sigma_{k}^{z}+\Delta_{m}\sigma_{m}^{z},\label{HI}
\end{align}
with $J=\sum_{n}g_{k,n}g_{m,n}(\mu^{2}-\omega_{n}^{2})^{-1}$ being the spin-spin coupling between the two ions.
By imposing the conditions \eqref{cond1} or \eqref{cond2} we realize the desired circulant Hamiltonian.
Note that such dispersive spin-phonon interaction was studied in the context of quantum simulation of effective spin models \cite{Kim2010} as well as for high-fidelity gate implementation \cite{Kim2009}.

\section{Numerical Examples}\label{ne}

Here we discuss specific time dependences of the detunings and the couplings which can be used to perform the gate implementation.
Consider first the \emph{Cases} \emph{1} and \emph{2} where the adiabatic transition to the quantum Fourier modes can be realized by using an exponential ramp of the detunings, $\Delta_{k}(t)=\Delta_{k}e^{-\gamma t}$ ($\Delta_{k}\gg J,\Omega_{1}$), with a characteristic rate $\gamma$.
Such a time dependence captures the asymptotic behaviour of the eigenvectors.
Another convenient choice of the time-dependent couplings and detunings, which we use for numerical examples, is
\begin{eqnarray}
&&J(t)=J_{0}\sin^{2}(\omega t),\quad \Omega_{1}(t)=\Omega_{1}\sin^{2}(\omega t),\notag\\
&&\Delta_{k}(t)=\Delta_{k}\cos^{2}(\omega t), \quad (k=1,2),\label{parameters}
\end{eqnarray}
where $\omega$ is a characteristic parameter which controls the adiabaticity of the transition.
The interaction time varies as $t\in[0,t_{\rm max}]$ with $t_{\rm max}=\pi/(2\omega)$.
This time dependence ensures that $\Delta_{k}(0)\gg J(0),\Omega_{1}(0)$, and respectively, $\Delta_{k}(t_{\rm tmax})\ll J(t_{\rm tmax}),\Omega_{1}(t_{\rm tmax})$.
\begin{figure}
\includegraphics[width=0.45\textwidth]{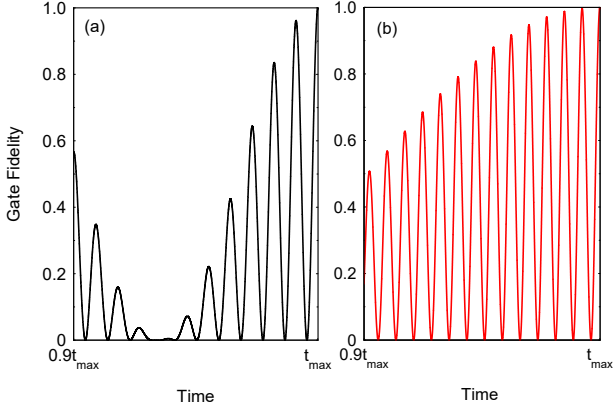}
\caption{(Color online) (a) Two-qubit fidelity calculated from the numerical simulation with Hamiltonian $\hat{H}(t)=\hat{H}_{0}+\hat{H}_{\rm cir}^{(2)}$.
The parameters are set to $\Omega_{1}/2\pi=40$ kHz, $\varphi=\pi/4$, $J_{0}/2\pi=2$ kHz and $\omega/2\pi=0.18$ kHz.
We choose the detunings $\Delta_{1}/2\pi=59.96$ kHz and $\Delta_{2}/2\pi=27.76$ kHz such that the adiabatic phases becomes $\alpha_{2}=2k\pi$, $\beta_{2}=2p\pi$ ($k=40$ and $p=20$) which realise the two qubit gate (\ref{gate}). (b) Fidelity of the adiabatic transition using Hamiltonian (\ref{H3}). The parameters are set to $J_{0}/2\pi=2$ kHz, $V_{0}/2\pi=2.02$ kHz, $\Omega_{1}/2\pi=146.3$ kHz, $\omega/2\pi=0.55$ kHz, and $\varphi=\pi/4$. }
\label{fig4}
\end{figure}

Finally, the adiabatic transition to the Fourier states using Hamiltonian (\ref{H3}) can be carried out by using $\Delta_{1,2}=0$,
\begin{equation}
J(t)=J_{0}\sin^{2}(\omega t),\quad \Omega_{2}(t)=J_{0}+V_{0}\cos^{2}(\omega t),\label{parameters_new}
\end{equation}
and $\Omega_{1}(t)=\Omega_{1}$. Again, initially we have $\Omega_{1},\Omega_{2}(0)\gg J(0)$ and respectively at the end of the transition $J(t_{\rm max})=\Omega_{2}(t_{\rm max})$ which ensures the circulant symmetry of the Hamiltonian (\ref{H3}).
\subsection{Eigenfrequencies}

In Fig. \ref{fig1} we plot the eigenfrequencies (\ref{ef}) and (\ref{ef1}) as a function of time.
We see that the eigenfrequencies for both cases are nondegenerate during the time evolution.
Approaching the final interaction time the energy separation between the adiabatic levels for the Hamiltonian (\ref{H1}) is determined by the coupling strength $J_{0}$, see Eq. (\ref{cr1}). For the circulant Hamiltonian (\ref{H2}) the separation between eigenfrequencies $\lambda_{\pm}$ and $\mu_{\pm}$ is again determined by $J_{0}$.
However, the presence of the single-qubit Rabi frequency $\Omega_{1}(t)$ leads to higher separation between the eigenfrequencies $\lambda_{+}$, $\mu_{+}$, and $\lambda_{-}$, $\mu_{-}$, where the energy gap is determined by $\Omega_{1}$ ($\Omega_{1}\gg J_{0}$), see Fig. \ref{fig1}(b).

\subsection{Gate fidelity}
\begin{figure}
\includegraphics[width=0.45\textwidth]{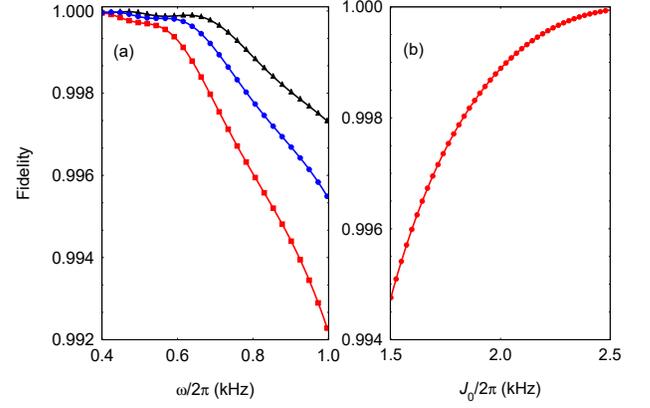}
\caption{(Color online) (a) Entangled state fidelity state calculated from the numerical simulation of Hamiltonian (\ref{H3}) as a function of $\omega$ for various $J_{0}$. The parameters are set to $\Omega_{1}/2\pi=30$ kHz, $V_{0}/2\pi=2.0$ kHz, $\varphi=\pi/4$, and $J_{0}/2\pi=2.0$ kHz (black triangles), $J_{0}/2\pi=1.8$ kHz (blue circles), $J_{0}/2\pi=1.8$ kHz (red squares). (b) The same but now set $\omega/2\pi=0.8$ kHz and vary the coupling strength $J_{0}$. }
\label{fig5}
\end{figure}

We numerically simulate the adiabatic transition to the quantum Fourier states (\ref{Fstates}) using the time-dependent couplings and detunings (\ref{parameters}) as well as (\ref{parameters_new}).
In Fig. \ref{fig2}(a) we plot the time evolution of the spin populations assuming that the system is prepared initially in the product state $\left|\psi(0)\right\rangle=\left|\downarrow\downarrow\right\rangle$.
We observe that even for the relatively small coupling $J_{0}$ the adiabatic transition transforms the initial state into the respective quantum Fourier state, namely $\left|\downarrow\downarrow\right\rangle\rightarrow\left|\psi_{3}\right\rangle$.
In this case the nonadiabatic transition is suppressed due to the single-qubit Rabi frequency $\Omega_{1}$ which improves the adiabaticity of the transition.
We have found that all other initial computational spin states approach the respective quantum Fourier states according to Eq. (\ref{transition}).
We also show the adiabatic transition $\left|--\right\rangle\rightarrow\left|\psi_{3}\right\rangle$ using Hamiltonian (\ref{H3}), see Fig. \ref{fig2}(b). We observe that compared to the Case 2 now the adiabatic transition is performed for shorter interaction time.

In Fig. \ref{fig3} we plot the time evolution of the arguments of the probability amplitudes for the different spin states.
The arguments tend toward the respective phases given by Eq. (\ref{Fstates}).
The same result also is observed for all other initial computational states.

As a figure of merit for the fidelity of the gate implementation we  use
\begin{equation}
F_{\rm gate}(t)=\tfrac{1}{16}|\sum_{s_{1},s_{2}}\langle s_{1}s_{2}|G^{\dag}_{\frac{\pi}{4}}G^{\prime}_{\frac{\pi}{4}}(t)|s_{1}s_{2}\rangle|^{2},\label{fidelity}
\end{equation}
where $s_{k}=\uparrow_{k},\downarrow_{k}$.
Here $G_{\frac{\pi}{4}}$ is the desired two-qubit quantum Fourier transform (\ref{gate}) and $G^{\prime}_{\frac{\pi}{4}}(t)$ is the actual one.
In Fig. \ref{fig4}(a) we show the two-qubit fidelity (\ref{fidelity}) as a function of time where we choose the detunings $\Delta_{1}$, $\Delta_{2}$ such that the adiabatic phases become $\alpha_{2}=2k\pi$, $\beta_{2}=2p\pi$.
As the time progresses the unitary propagator $G_{\frac{\pi}{4}}^{\prime}$ converges toward $G_{\frac{\pi}{4}}$.
We observe that for spin-spin coupling $J_{0}/2\pi=2$ kHz and gate time $t_{\rm max}\approx 1.4$ ms one can achieve gate infidelity of $1-F_{\rm gate}(t_{\rm max})\approx 10^{-4}$. In Fig. \ref{fig4}(b) we plot the fidelity of the adiabatic transition between the rotating computational spin states $\left|q_{1},q_{2}\right\rangle$, ($q_{k}=\pm_{k}$) and the quantum Fourier states (\ref{Fstates}), using Hamiltonian (\ref{H3}) (see the Supplement for more details). We observe high fidelity of the adiabatic transition within shorter interaction time $t_{\rm max}\approx 455$ $\mu$s.

\subsection{Creation of entangled states}

The action of the two-qubit gate on the computational basis creates superposition states which, however, are not entangled.
In order to create entangle states one needs to prepare initially the system is a superposition spin state.
For example, consider that the initial state is $\left|\psi(0)\right\rangle=(e^{-i\alpha_{2}}\left|\downarrow_{1}\right\rangle+e^{i\beta_{2}}\left|\uparrow_{1}\right\rangle)\left|\downarrow_{2}\right\rangle/\sqrt{2}$.
The two-qubit gate (\ref{transition}) transforms the initial state into an entangled state which is superposition of two Fourier modes, namely $\left|\psi(0)\right\rangle\rightarrow\left|\psi(t_{\rm f})\right\rangle=(\left|\psi_{3}\right\rangle+\left|\psi_{2}\right\rangle)/\sqrt{2}$. The same state can be created also by preparing initially the system in the rotating superposition state $\left|\psi_{\rm r}(0)\right\rangle=(e^{-i\alpha}\left|-_{1}\right\rangle+e^{-i\beta}\left|+_{1}\right\rangle)\left|-_{2}\right\rangle/\sqrt{2}$. Then adiabatically following the instantaneous eigenstates of Hamiltonian (\ref{H3}) one can perform the transition $\left|\psi_{\rm r}(0)\right\rangle\rightarrow\left|\psi(t_{\rm f})\right\rangle$.
In Fig. \ref{fig5} we show the fidelity of the creation of the entangled state defined by $F(t)=\frac{1}{2}|\langle\psi(t_{\rm f})|(e^{-i\alpha}\left|\chi_{-}(t)\right\rangle+e^{-i\beta}\left|\nu_{-}(t)\right\rangle)|^{2}$ as a function of $\omega$ and $J_{0}$, where $\left|\chi_{-}(t)\right\rangle$ and $\left|\nu_{-}(t)\right\rangle$ are the instantaneous eigenstates (see the Supplement for the derivation).
As can be seen by lowering $\omega$ the adiabaticity of the transition is improved which leads to higher fidelity.
For example, for $J_{0}/2\pi=2$ kHz and $\omega/2\pi=0.8$ kHz with gate time $t_{\rm max}=313$ $\mu$s we estimate infidelity of order of $1-F(t_{\rm max})\approx 10^{-4}$.

\section{Shortcut to adiabaticity}

\begin{figure}
\includegraphics[width=0.45\textwidth]{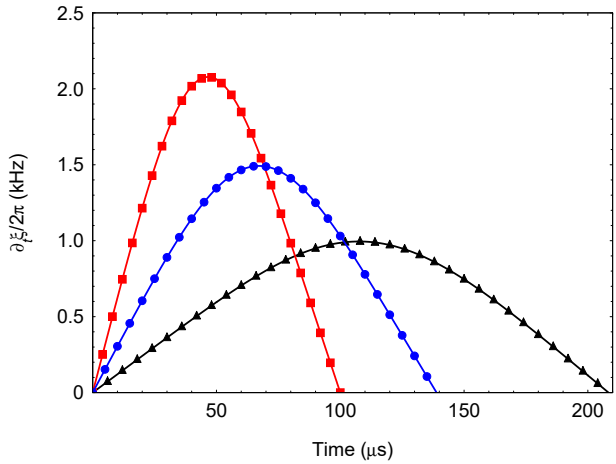}
\caption{(Color online) Shape of the counterdriving field (\ref{ps}) as a function of time for various values of $\omega$ and $J_{0}$.
We set $\Omega_{1}/2\pi=80$ kHz, $V_{0}/2\pi=0.5$ kHz. The other parameters are $\omega/2\pi=2.5$ kHz, $J_{0}/2\pi=2.0$ kHz (red squares), $\omega/2\pi=1.8$ kHz, $J_{0}/2\pi=1.5$ kHz (blue circles), $\omega/2\pi=1.2$ kHz, $J_{0}/2\pi=1.0$ kHz (black triangles).}
\label{fig6}
\end{figure}

Finally we discuss the possibility to apply a counterdriving field which suppresses the nonadiabatic transitions.
For concreteness we focus on the implementation using the Hamiltonian \eqref{H3} because it allows us to derive an explicit analytic expression for the instantaneous eigenstates.
Following \cite{Odelin2019} the total Hamiltonian including the counterdriving field becomes
\begin{equation}
\hat{H}_{\rm T}=\hat{H}+\hat{H}_{\rm{CD}},\quad \hat{H}_{\rm CD}=i\sum_{s=\pm}\{|\partial_{t}\chi_{s}\rangle\langle\chi_{s}|+|\partial_{t}\nu_{s}\rangle\langle\nu_{s}|\},
\end{equation}
where the second term cancels the nonadiabatic coupling.
Here $\left|\chi_{\pm}(t)\right\rangle$ and  $\left|\nu_{\pm}(t)\right\rangle$ are the time-dependent eigenstates of (\ref{H3}). We find
\begin{equation}
\hat{H}_{\rm CD}=-\partial_{t}\xi\{\left|\downarrow_{1}\right\rangle\left\langle\uparrow_{1}\right|+
\left|\uparrow_{1}\right\rangle\left\langle\downarrow_{1}\right|\}\left|\downarrow_{2}\right\rangle\left\langle\downarrow_{2}\right|,
\end{equation}
where the mixing angle is $\tan(\xi)=\Omega_{2}/J$.
Using the time-dependent couplings (\ref{parameters_new}) we obtain
\begin{equation}\label{ps}
\partial_{t}\xi=\frac{\omega J_{0}(J_{0}+V_{0})\sin(2\omega t)}{J_{0}^{2}\sin^{4}(\omega t)+[V_{0}\sin^{2}(\omega t)-(J_{0}+V_{0})]^2}.
\end{equation}

In Fig. \ref{fig6} we show the shape of the counterdriving field (\ref{ps}) for various values of $\omega$ and $J_{0}$. We see that the countrerdriving field vanishes at $t=0$ which preserves the requirement system to begin in the rotating spin states. At $t_{\rm max}$ we have $\partial_{t}\xi(t_{\rm max})$ such that the system end up in state with circulant symmetry.
Importantly, we observe that for the same magnitude of $J_{0}\sim \partial_{t}\xi$ one can reduce the gate time such that $\omega>J_{0}$.
Consider as an example spin coupling $J_{0}/2\pi=2.0$ kHz.
For approximately the same maximal magnitude of $\partial_{t}\xi$ the gate time is approximately a factor of four shorter, $\omega/2\pi=2.5$ kHz and $t_{\rm max}=100$ $\mu$s, see Fig. \ref{fig4}(b) for comparison.

\section{Conclusion}\label{C}

We have shown that using a Hamiltonian with the circulant symmetry one can realize the two-qubit quantum Fourier transform.
The unique property of the circulant Hamiltonian is that its eigenvectors are the quantum Fourier modes.
Our model consists of two interaction spins which are subjected to an additional single-qubit drive.
We have considered the conditions for the spin coupling and the single-qubit Rabi frequencies which lead to the circulant symmetry of the spin Hamiltonian.
Our two-qubit gate scheme is based on an adiabatic transition of the computational spin basis into the respective quantum Fourier modes which realizes the quantum Fourier transform in a single interaction step.
We have discussed the physical implementation of the circulant Hamiltonian using trapped ions.
The realization relies on using a bichromatic laser field which couples the internal ion's states with the collective vibrational modes.
We discuss the fidelity of the gate operation as well as the fidelity of the entangled-state creation.
We have shown that the actual two-qubit gate converges with infidelity of order of $10^{-4}$ toward the desired quantum Fourier transform.
Finally, we described how the gate can be accelerated by at least a factor of 4 by using  a counterdiabatic shortcut.

%%%%%%%%%%%%%%%%%%%%%%%
%%%%%%%%%%%%%%%%%%%%%%%%%%%%%%%%%
%%%%%%%%%%%%%%%%%%%%%%%%%%%%%%%%%
%%%%%%%%%%%%%%%%%%%%%%%%%%%%%%%%%
%%%%%%%%%%%%%%%%%%%%%%%%%%%%%%%%%
%%%%%%%%%%%%%%%%%%%%%%%%%%%%%%%%%
%%%%%%%%%%%%%%%%%%%%%%%%%%%%%%%%%
%%%%%%%%%%%%%%%%%%%%%%%%%%%%%%%%%%%%%%%%%%%%%%%%%%%%%%%%%%%%%%%%%%%%%%%%%%%%%%%%%%%%%%%%%%%%%%%%%%%%%%
%%%%%%%%%%%%%%%%%%%%%%%%%%%%%%%%%%%%%%%%%%%%%%%%%%%%%%%%%%%%%%%%%%%%%%%%%%%%%%%%%%%%%%%%%%%%%%%%%%%%%%

\section*{Acknowledgments}
PAI acknowledges suport from the European Commission's Horizon-2020 Flagship on Quantum Technologies project 820314 (MicroQC).
NVV acknowledges support from the Bulgarian Science Fund Grant DO02/3 (Quant-ERA project ERyQSenS).

\section*{Author Contributions}

P. A. I. and N. V. V developed the concept. The main calculations and numerical simulations were performed by P. A. I. Both authors contributed to the analysis of the results. Both authors wrote the manuscript.

\section*{Additional Information}
\textbf{Competing financial interests}: The authors declare no competing financial and/or nonfinacial interests.

\end{document}